# Quantum Critical Behavior in the Asymptotic Limit of High Disorder: Entropy Stabilized NiCoCr$_{0.8}$ Alloys

Brian C. Sales[1]*, Ke Jin[1], Hongbin Bei[1], John Nichols[1], Matthew F. Chisholm[1], Andrew F. May[1], Nicholas P. Butch[2,3], Andrew D. Christianson[4] and Michael A. McGuire[1]

[1]Materials Science and Technology Division, Oak Ridge National Laboratory, Oak Ridge TN
[2]NIST Center for Neutron Research, National Institute of Standards and Technology, Gaithersburg, Maryland
[3]Center for Nanophysics and Advanced Materials, Department of Physics, University of Maryland, College Park, Maryland
[4]Quantum Condensed Matter Division, Oak Ridge National Laboratory, Oak Ridge TN

**\*Correspondence to:  salesbc@ornl.gov**

## Abstract

The behavior of matter near a quantum critical point (QCP) is one of the most exciting and challenging areas of physics research. Emergent phenomena such as high-temperature superconductivity are linked to the proximity to a QCP. Although significant progress has been made in understanding quantum critical behavior in some low dimensional magnetic insulators, the situation in metallic systems is much less clear. Here we demonstrate that NiCoCr$_x$ single crystal alloys are remarkable model systems for investigating QCP physics in a metallic environment. For NiCoCr$_x$ alloys with x ≈ 0.8, the critical exponents associated with a ferromagnetic quantum critical point (FQCP) are experimentally determined from low temperature magnetization and heat capacity measurements. For the first time, all of the five exponents ( $\gamma_T \approx 1/2$ , $\beta_T \approx 1$, $\delta \approx 3/2$, $\nu z_m \approx 2$, $\overline{\alpha}_T \approx 0$)  are in remarkable agreement with predictions of Belitz-Kirkpatrick-Vojta (BKV) theory in the asymptotic limit of high disorder. Using these critical exponents, excellent scaling of the magnetization data is demonstrated with no adjustable parameters. We also find a divergence of the magnetic Gruneisen parameter, consistent with a FQCP. This work therefore demonstrates that entropy stabilized concentrated solid solutions represent a unique platform to study quantum critical behavior in a highly tunable class of materials.

## Introduction

The connection between quantum fluctuations in the vicinity of a quantum critical point (QCP) and emergent ground states, such as high-temperature superconductivity, remains a topic of great interest in the condensed matter physics community.[1-5] One key to making progress is this area is the identification of model material systems that are complex enough to exhibit the physics of interest but simple enough to be compared to theory. Studies of several low dimensional magnetic materials have greatly improved our understanding of QCP physics in insulating solids,[6-8] but our understanding of QCP phenomena in metallic systems is much less clear.[2, 9,10] One hindrance is that few systems are microscopically homogeneous near the QCP, due to the need for small concentrations of dopants, and this can disrupt the feedback loop between theory and experiment.

The $NiCoCr_x$ alloys (with $x \approx 1$) are related to the recently discovered high-entropy-alloys, such as $NiCoCrFeMn$[11-12], where configurational entropy stabilizes a random distribution of elements on a simple face-centered cubic lattice (fcc). These alloys have been shown to be chemically homogeneous from the centimeter to the nanometer scale. An example of an atomic resolution scanning transmission electron microscope (STEM) image from a NiCoCr single crystal is shown in Figure 1. No clustering of any of the three elements is observed.

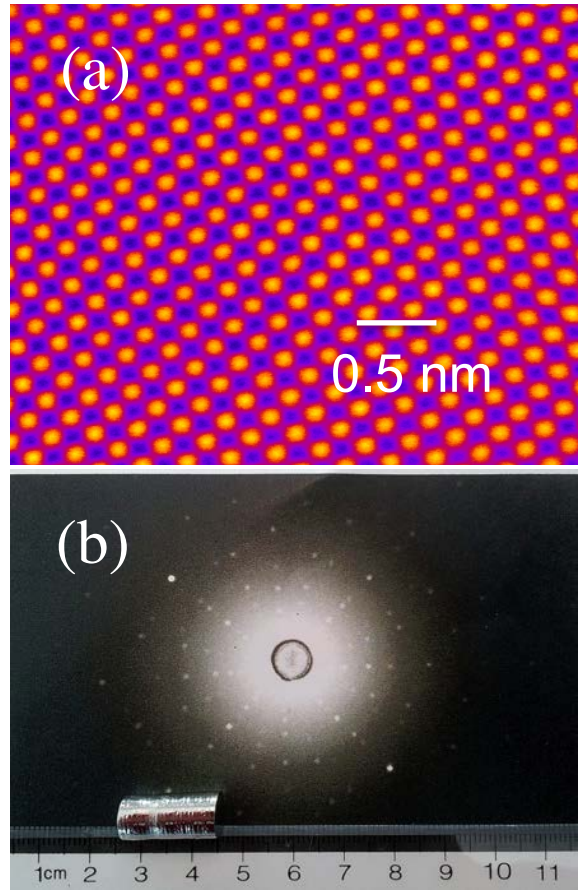

Fig. 1. **Homogeneous NiCoCr Crystals** (a) STEM image of columns of atoms along the [100] direction from a NiCoCr single crystal. These images and associated EELS (electron energy loss spectroscopy) data taken at the same time do not detect any evidence of elemental clustering or ordering. Each "atom" in the figure is actually a column of atoms about 20 nm in length. The distance between columns is approximately 0.18 nm. The images were obtained using a Nion UltraSTEM200 operated at 200 kV with an illumination half-angle of 30 mrad and an inner detector half-angle of 65 mrad. (b) Laue x-ray image from a 10 g NiCoCr$_{0.8}$ crystal shown at the bottom.

The high-entropy-alloys have remarkable mechanical properties[13], and enhanced resistance to radiation damage.[14] The alloys are also unusual because all elements have roughly equal concentrations and the dilute impurity approximation is invalid. A recent theoretical model based on enthalpies of formation coupled with entropy stabilization successfully predicts which combination of elements will form a single-phase solid solution.[15] The composition NiCoCr also has remarkable mechanical properties at cryogenic temperatures and has been termed a medium-entropy alloy.[16]

Here we report low temperature magnetization and heat capacity for single crystals of the concentrated solid solutions $NiCoCr_x$ with $x \approx 0.8$ as a function of temperature and magnetic field. The extreme chemical disorder in these alloys is an essential component of the physics, suppressing a first order ferromagnetic transition that invariably occurs in many other potential FQCP systems.[9] In carefully prepared single crystals with electropolished surfaces and minimum mechanical damage, we find that the BKV theory[17,21] of metallic ferromagnetic quantum critical points provides an excellent description of all of our data. The approximate temperature–magnetic field phase diagram that describes the boundary between the quantum critical and the Fermi liquid regions is also determined for the $NiCoCr_{0.8}$ composition from magnetization data. Finally we demonstrate that the magnetic Gruneisen parameter $[\Gamma = (dM/dT)/ C_p]$ diverges, another indication of a FQCP. These findings open the door to the study of QCP physics in a new and flexible class of model materials.

**Results**

As is illustrated in Fig. 2, $NiCoCr_{0.8}$ is close to the Cr concentration where the ferromagnetic transition temperature, $T_c$, goes to zero. The ferromagnetism is highly itinerant for all compositions $x > 0.5$; for values of $x \approx 0.6$ the saturation moment per atom is 0.15 $\mu_B$, $T_c = 75 \pm 2$ K, and the Rhodes-Wohlfarth ratio is 8.5.[18] For increasing Cr concentrations, x, there is a rapid nearly exponential decrease in $T_c$, and in addition to the rapid depression of $T_c$, increasing the Cr concentration introduces a type of frustration since the spins on the Cr atoms want to be antiparallel to neighboring Ni, Co and Cr spins, which cannot be satisfied on a fcc lattice. For Cr concentrations greater than $x \approx 0.4$, theory greatly over-estimates both the spontaneous magnetic moments and $T_c$ (see Fig 3c. Ref. 18) indicating significant effects that are not captured by density functional theory. This deviation from theory and the observation of $T_c \to 0$ K motivate a detailed characterization of the critical behavior and exponents for compositions near $NiCoCr_{0.8}$.

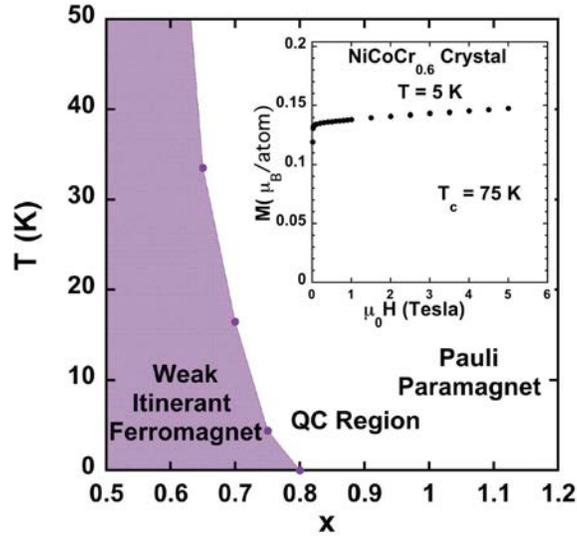

Fig. 2. **Phase diagram of NiCoCr$_x$ alloys.** Illustration of the rapid non-linear decrease in Curie temperature, $T_c$, with chromium concentration, x. The Curie temperatures for each composition are estimated from Arrott plots, the maximum in -dM/dT with H= 100 Oe, or extrapolation of $M^2$ vs T with H = 100 Oe. The $T_c$'s determined by each method are self-consistent within an error of about ± 2 K. (see ref 18 for additional details). The inset shows a magnetization curve for a NiCoCr$_{0.6}$ crystal at 5 K. The small value of the spontaneous moment indicates highly itinerant ferromagnetism.

The low field and low temperature magnetization data from a NiCoCr$_{0.8}$ single crystal is shown in Fig 3a. The divergence of the low field susceptibility, $\chi$= (M/H) as T approaches zero determines the critical exponent $\gamma_T$, where $\chi \propto T^{-\gamma_T}$. For low applied magnetic fields from 0.0001 to 0.1 Tesla, the susceptibility of NiCoCr$_{0.8}$ crystals diverge with $\gamma_T \approx 0.5$. It is important to note that to observe this behavior the surface of the crystals had to be electropolished and carefully cleaned before magnetic measurements (see experimental section), otherwise the magnetic response is dominated by extrinsic defects.

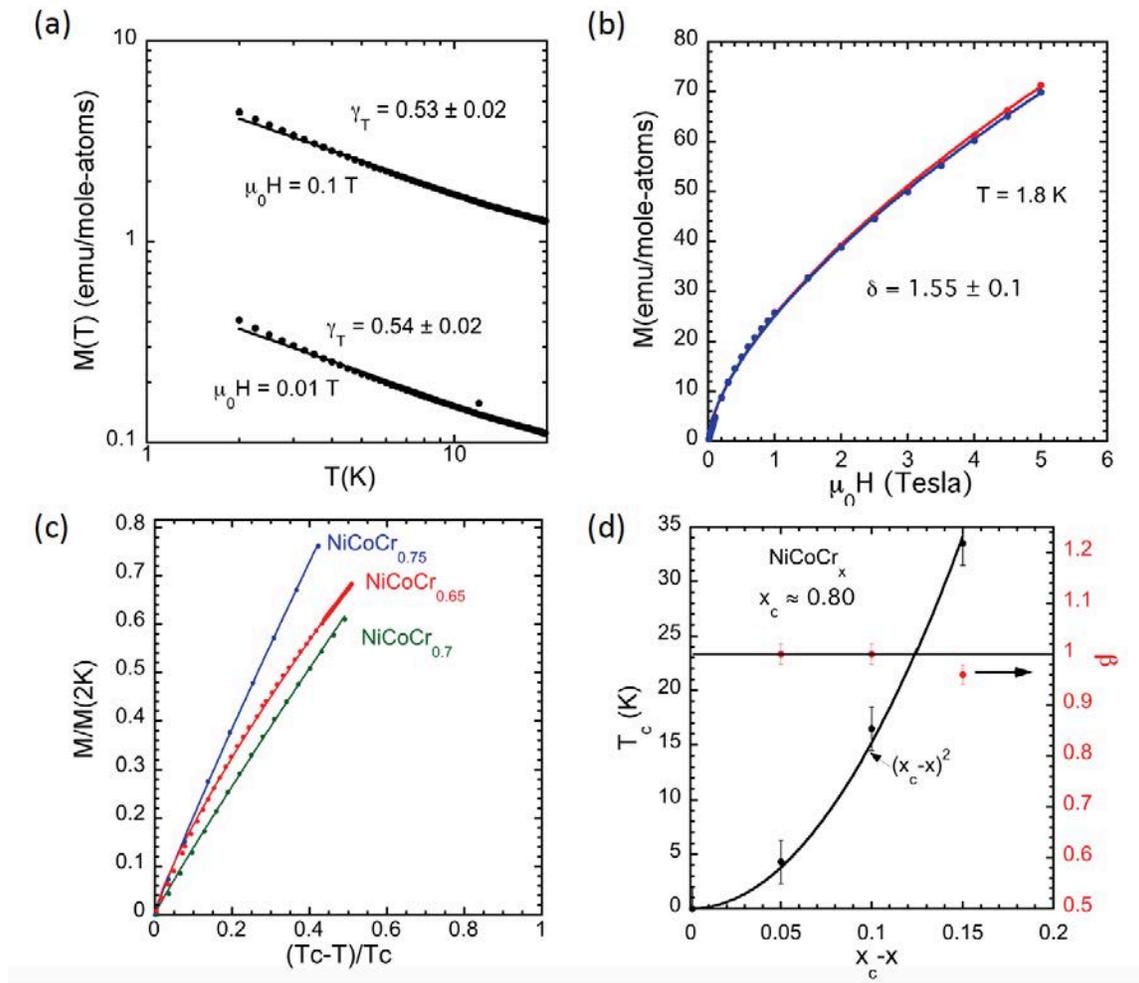

Fig. 3. **Critical Exponents from magnetization data** (a) Magnetization vs temperature for a NiCoCr$_{0.8}$ crystal in applied magnetic fields of $\mu_0 H$ = 0.01 and 0.1 Tesla. For these low applied fields the susceptibility, $\chi$, is given by M/H. The solid lines are least square fits of the data to a power law $T^{-\gamma_T}$, with the values for $\gamma_T$ as shown in the figure. (b) magnetization vs applied magnetic field at our lowest temperature for two NiCoCr$_{0.8}$ crystals, and the value of $\delta$ determined from these isotherms (c) Increase of the magnetization below Tc with an applied field of 0.01 T for three ferromagnetic polycrystalline samples. The lines are power law fits to all the data shown. Values of $\beta$ are obtained by fits over smaller T$_c$-T intervals near T$_c$. (d) Curie temperature versus x$_c$-x, where x is the Cr concentration, and x$_c$ is the concentration where T$_c \approx$ 0. The solid line is a force fit of the T$_c$ data to (x$_c$ - x)$^2$, the functional dependence expected from BKV theory. The values of the critical exponent $\beta$ as estimated for the same ferromagnetic samples x= 0.75, 0.7, 0.65. (1 emu = 1mA m$^2$ in SI units). Error bars correspond to one standard deviation.

The critical exponent, δ, is determined from the critical isotherm in the limit that T $\rightarrow$ 0 K and is defined as M(T,H) α H$^{1/\delta}$. Although technically δ is only defined at T$_c$ (which is 0 at the FQCP), the values are expected to saturate at a low enough temperature. Magnetization isotherms at our lowest temperature of 1.8 K are displayed in Fig. 3b for two NiCoCr$_{0.8}$ crystals. Power law fits to these data yield an estimate of δ = 1.55 ± 0.1.

The critical exponent, β$_T$, can be formally associated with a FQCP and is defined in terms of the order parameter M in the limit of H $\rightarrow$ 0 and T $\rightarrow$ 0 as proportional to T$^{\beta_T}$. The proportionality constant is of course zero at the FQCP since the spontaneous magnetization is only non-zero in the ordered magnetic phase. In this case the value of β$_T$ can be estimated by tuning the Cr composition slightly away from the critical value of x ≈ 0.8 and into the ferromagnetic state (x < 0.8, see Fig. 2) and then determine the values of β as x =0.8 is approached. This was done for 3 polycrystalline NiCoCr$_x$ samples with x = 0.75, 0.70, 0.65 and T$_c$'s of 4.4, 16.5 and 33.5 K. These values for T$_c$ were defined by the maximum in -dM/dT with H =100 Oe. This definition of T$_c$ appeared to be the least subject to assumptions. We note however that changing T$_c$ by a couple of degrees did not significantly effect the results for β. The evolution of the magnetization below T$_c$ is evaluated for each of the three alloys with a small applied field of H=100 Oe. For each alloy the magnetization data below T$_c$ was fit to a power law for various values of T$_c$-T, with the assumption that the correct value of β is attained as T$_c$-T approaches zero (Fig 3c). Using this analysis, we obtain the value of β for each alloy. These values are plotted in Fig 3d and yield an approximate value for β$_T$ (x ≈ 0.8) of 1.

Taken together, direct analysis of the magnetization data for NiCoCr$_{0.8}$ give the following critical exponents: γ$_T$ ≈ 0.5, δ ≈ 3/2 , and β$_T$ ≈ 1. The subscript T for γ and β simply indicate a quantum critical point where the transition temperature is zero.

This is also the notation used in the BKV theory[17,21]. These experimental values for $\gamma_T$, $\delta$, and $\beta_T$ are also the values predicted from BKV theory in the "dirty" or high-disordered limit. In this limit the Widom relationship $\gamma_T = \beta_T (\delta - 1)$, is expected to hold[21] and within our experimental error it does. This means that only two of the three exponents are independent. In the "dirty" limit, BKV theory[21] also predicts that the Curie temperature should increase rapidly as the composition is tuned away from the critical composition as $T_c \propto (x_c-x)^{z_m\nu} = (x_c - x)^2$, since $z_m = 2$ and $\nu = 1$ (from the theory). Although we do not have enough experimental data to show that $z_m\nu$ is exactly 2, a force fit of the $T_c$ data to $(x_c - x)^2$ does provide a good description of our data as shown in Fig. 3d.

One test of quantum critical behavior is the divergence of the magnetic Gruneisen parameter[22] (also called the magnetocaloric ratio), $\Gamma = (dM/dT)/C_p$, as T approaches 0.[10,21,22] Low temperature heat capacity data from $NiCoCr_{0.8}$ with H=0 are displayed in Fig 4a. At the lowest temperatures $C_p$ is proportional to $T^\alpha$ with a value of $\alpha = 0.92$, which implies that $C_p/T$ weakly diverges as $T^{-0.08}$. This weak divergence of $C_p/T$ is close to the exponent value of $\overline{\alpha}_T = 0$ expected from BKV theory. As noted above, the low field (H=100 Oe) susceptibility (M/H), diverges as $\approx T^{-0.5}$, which implies that $dM/dT$ diverges as $\approx T^{-3/2}$. Thus $\Gamma$ diverges as $\approx T^{-2.5}$. This low temperature divergence is demonstrated in Fig 4b, using either the total measured heat capacity or just the electronic/magnetic portion.

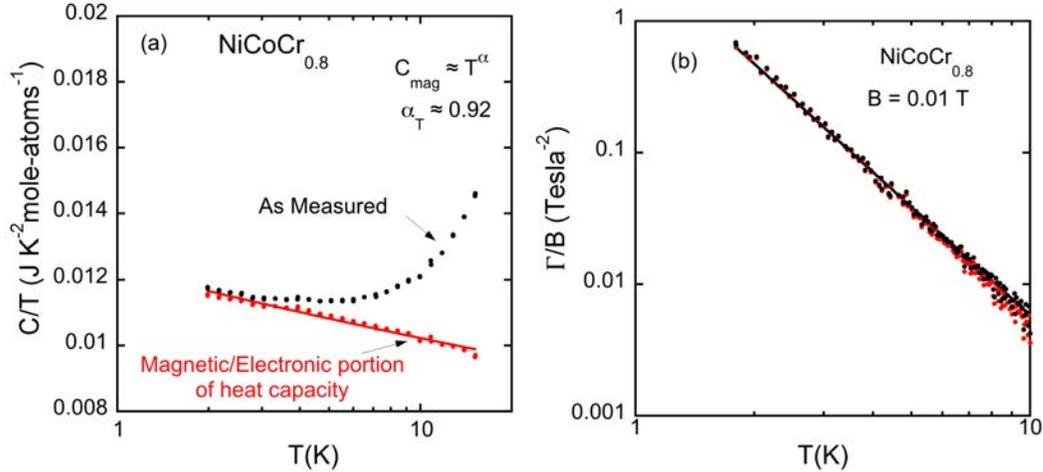

Fig. 4. **Heat capacity data and illustration of divergence of magnetic Gruneisen coefficient** (a) Heat capacity data divided by temperature vs log T. The lattice contribution is estimated using a NiCoCr$_{1.2}$ alloy and it is subtracted from the total measured heat capacity for NiCoCr$_{0.8}$ resulting in the data shown in red. This alloy is a Pauli paramagnet outside of the critical region (see Fig 2.) that should have a very similar phonon contribution to the heat capacity.[18] The low temperature heat capacity of the reference alloy, NiCoCr$_{1.2}$, is well described by a Debye model with an electronic contribution characterized by $\gamma$= 9.2 mJ/mole-K$^2$, and $\Theta_D$ = 466 K.  (b) log ($\Gamma$) versus log (T) with a solid line illustrating the divergence of  $\Gamma \approx$  T$^{-2.5}$.

The scaling of the magnetization data, M(T, H), are determined by the critical exponents[19,20]   All of the magnetization data from the NiCoCr$_{0.8}$ single crystals should collapse onto a single curve when M/T$^{\beta_T}$ is plotted versus H/T$^{\gamma_T + \beta_T}$.  Since $\gamma_T$ and $\beta_T$ for NiCoCr$_{0.8}$ are found experimentally to have the values of $\gamma_T \approx$ 0.5, and $\beta_T$ $\approx$ 1, this implies that if  M(T, H)/T is plotted versus H/T$^{1.5}$, all of the magnetization data should fall on one curve. This is indeed the case as illustrated in Fig 5. It is important to emphasize that we are not using the scaling plots to extract precise values for $\gamma_T$ and $\beta_T$, but rather we simply illustrate that the values found do result in excellent scaling. It has been demonstrated recently that it is very difficult to extract unique values of critical exponents from scaling plots alone.[19]

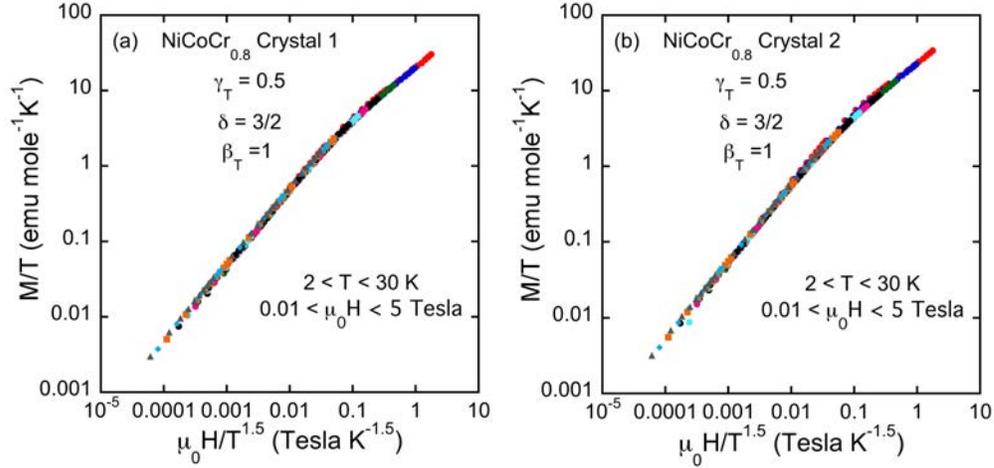

Fig. 5. **Demonstration of excellent scaling** of magnetization data from two crystals (a,b) of NiCoCr$_{0.8}$ using the experimentally determined values of of $\gamma_T \approx 0.5$, and $\beta_T \approx 1$. There are no adjustable parameters.

## Discussion

We have provided strong evidence that NiCoCr0.$_8$ is close to a FQCP and the experimentally determine critical exponents ( $\gamma_T \approx 0.53 \pm 0.02$ , $\beta_T \approx 1 \pm 0.02$, $\delta \approx 1.55 \pm 0.1$, $\nu z_m = 2$, $\overline{\alpha}_T = 0.08$) are close to the values expected from BKV theory in the asymptotic limit of high-disorder ( $\gamma_T = 1/2$ , $\beta_T = 1$, $\delta = 3/2$, $\nu z_m = 2$, $\overline{\alpha}_T = 0.0$). Magnetic field, which can be thought of as another variable perpendicular to the temperature and Cr composition axis (Fig. 2), is used to probe the extent of the quantum critical region away from the FQCP of the NiCoCr$_{0.8}$ alloy. As a magnetic field is applied to NiCoCr$_{0.8}$ the magnetic fluctuations are suppressed and the system is tuned toward a normal Fermi liquid with a Pauli susceptibility. This behavior suggests there is a region where the system's behavior is governed by quantum fluctuations, as opposed to thermally-induced fluctuations, consistent with the existence of quantum critical behavior in NiCoCr$_{0.8}$. This crossover from the quantum critical regime to the Fermi liquid regime occurs at T* for a given applied magnetic field B. Experimentally the crossover temperature T* is usually estimated as the temperature where a maximum occurs in $\chi$(T), with $\chi$ =dM/dH extracted from DC magnetization curves at fixed temperatures or from ac susceptibility data[10,23] such as shown in Fig 6a. The dependence of T* on applied magnetic field B is shown in Fig 6b. A power law fit to T* vs B gives T* = 5.62 $B^{0.67}$ ( or $T^{*1.5}$ vs B) the same scaling variable used in Fig. 5. This provides another self-consistency check that the analysis of the magnetization data from NiCoCr$_{0.8}$ in terms of a FQCP is reasonable. As briefly discussed in the experimental section, strain can also be used as another tuning variable, though we have not developed a strain-based phase diagram to complement Fig. 6b.

Since there are well over two hundred ternary concentrated solid solutions with compositions ABC predicted to form,[25,15] this class of entropy stabilized alloys represents an interesting new class of tunable materials for the investigation of quantum critical behavior.

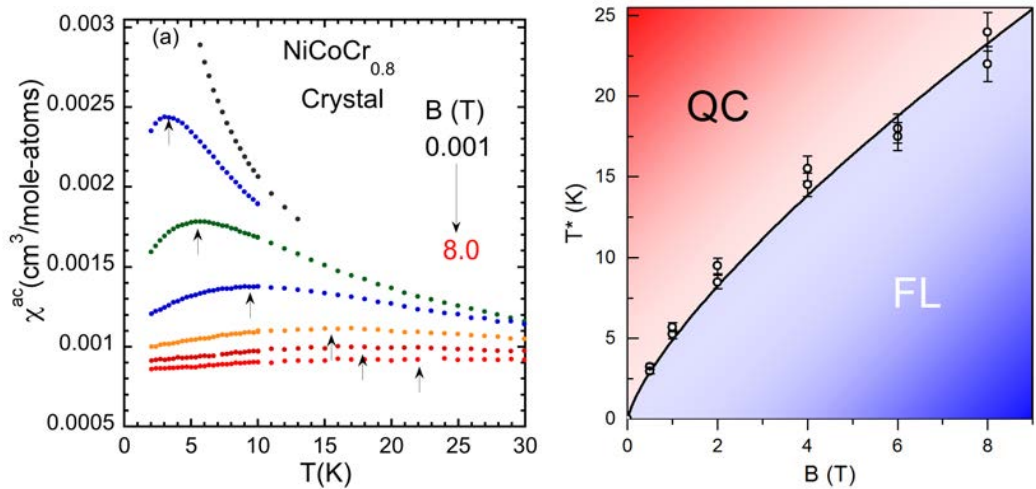

Fig. 6. **Field induced crossover from quantum critical to Fermi liquid regimes** (a) ac susceptibility versus temperature with applied dc magnetic fields between 0.001 and 8 Tesla. The black arrows denote the approximate values of T* for a fixed dc magnetic field. (b) The crossover temperature, T*, as a function of magnetic field for a NiCoCr$_{0.8}$ single crystal as determined from both DC magnetization and ac susceptibility data. The regions dominated by quantum critical (QC) and Fermi liquid (FL) behavior are noted in the figure. The black line is a power law fit to the data: T*= 5.62 $B^{0.67}$. Error bars correspond to one standard error.

**Methods**

Single crystals of $NiCoCr_{0.8}$ are grown in an optical floating zone furnace. Polycrystalline rods for the crystal growth are produced by arc-melting stoichiometric amounts (1:1:0.8) of high purity (> 99.9 %) Ni, Co, and Cr in an argon atmosphere followed by drop casting into a copper mold.[24] Laue back reflection and x-ray diffraction measurements indicated single crystals with a small mosaic spread ( $\approx$ 0.4°). Crystals suitable in size for magnetization ($\approx$200 mg) or heat capacity ($\approx$25 mg) are cut from a large single crystal boule using an electro-discharge machine (EDM). To eliminate the damage layer created after cutting or sanding, the crystals are electropolished in an 85% $H_3PO_4$ solution with approximately a 10 V bias. This surface preparation is essential, particularly for the magnetic measurements. If, for example, the samples are cut with a low speed diamond saw and then polished with SiC sandpaper, or just cut with a diamond saw, it is found that the temperature dependence and magnitude of the low field magnetic susceptibility is dramatically altered- by more than a factor of 2. We also find that severe coldworking the material increases the low temperature susceptibility by about a factor of 10. These results suggest that strain and pressure may also be a useful QCP tuning parameter for the $NiCoCr_{0.8}$ alloys, although that is beyond the scope of the present work. The chemical homogeneity of crystals grown this way have been checked on the micron scale using energy dispersive spectroscopy measurements and on the nanoscale using atom probe measurements.[24] To further examine both the crystalline perfection and the chemical homogeneity we perform atomic resolution scanning transmission electron microscopy (STEM) measurements and single column electron energy loss spectroscopy (EELS) on a NiCoCr single crystal. The STEM results, shown in Fig. 1, confirm the crystalline perfection and the absence of clustering of any of the elements.

The DC and AC magnetic measurements are made using a commercial SQUID magnetometer (MPMS from Quantum design) and the ac susceptibility option of a PPMS (also from Quantum Design), respectively. To check for possible glassy magnetic behavior, both zero-field-cooled and field-cooled measurements are performed. Within experimental error there is no difference in the magnetization data at all fields. As a further check, ac susceptibility measurements are made as a function of frequency, and no glassy behavior could be detected (Fig S1). Heat capacity measurements are made using the PPMS heat capacity option. Identification of commercial equipment does not imply endorsement by NIST.

**Acknowledgements**

It is a pleasure to acknowledge helpful discussions with Jamie Morris, Lekh Poudel, Malcolm Stocks, German Samolyuk, and Jiaqiang Yan. This research was supported primarily by the Department of Energy, Office of Science, Basic Energy Sciences, Materials Sciences and Engineering Division (B. C. S., J. N., A. F. M., M. F. C., M. A. M.). K. J. and H. B. were supported by the Energy Dissipation to Defect Evolution (EDDE), an Energy Frontier Research Center funded by the U. S. Department of Energy, Office of Science, BES. Research at ORNL's Spallation Neutron Source was supported by the Scientific User Facilities Division, Office of Basic Energy Sciences, U S. Department of Energy (A. D.)

**Author Contributions**

B.C.S. conceived the experiments, and made most of the transport, magnetic, and heat capacity measurements with help from A.F.M., M.A.M and J. N. K. J.  and H. B. grew, oriented, cut, and electropolished the crystals. M.F.C. took and analyzed atomic resolution STEM and EELS data. AC and NB made key contributions to the analysis of the data. All authors contributed to the preparation of the manuscript.

**Supplementary Information**
Supplementary information is attached

Supplementary Information

# Quantum critical behavior in NiCoCr$_x$ alloys: a new twist on structural materials

Brian C. Sales[1]*, Ke Jin[1], Hongbin Bei[1], John Nichols[1], Matthew F. Chisholm[1], Andrew F. May[1], Nicholas P. Butch[2,3], Andrew D. Christianson[4] and Michael A. McGuire[1]

[1]Materials Science and Technology Division, Oak Ridge National Laboratory, Oak Ridge TN
[2]NIST Center for Neutron Research, National Institute of Standards and Technology, Gaithersburg, Maryland
[3]Center for Nanophysics and Advanced Materials, Department of Physics, University of Maryland, College Park, Maryland
[4]Quantum Condensed Matter Division, Oak Ridge National Laboratory, Oak Ridge TN

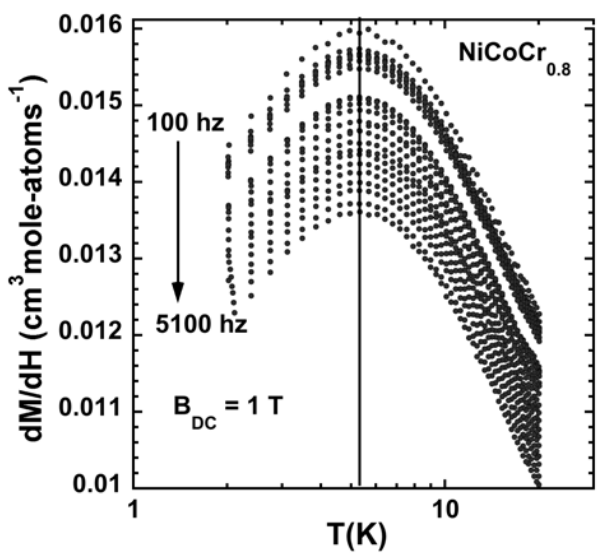

Figure S1 **No evidence of glassy behavior.** Real part of ac susceptibility of a NiCoCr$_{0.8}$ single crystal versus temperature for measuring frequencies between 100 and 5100 hz. For most glassy magnetic systems the position of the peak changes as a function of measuring frequency. Within experimental error, the peak position is unchanged for $NiCoCr_{0.8}$.